      [1999/12/01 v1.4c Il Nuovo Cimento]
\documentclass{cimento}

\usepackage{graphicx}  % got figures? uncomment this
\title{Present and Future Prospects for GRB Standard Candles}
\author{A.S.~Friedman\from{ins:x}\ETC,
\atque J.S.~Bloom\from{ins:x},\from{ins:y}}
\instlist{\inst{ins:x} Harvard Smithsonian Center for Astrophysics, 
	60 Garden Street, Cambridge, MA 02138.  \inst{ins:y}
  Dept. of Astronomy, 601 Campbell Hall, Univ. of California, Berkeley, Berkeley, CA 94720.}
\PACSes{\PACSit{95.85Pw},\PACSit{98.70Rz},\PACSit{98.80.-k},\PACSit{98.80.Es}{\ldots}}
\begin{document}

\maketitle

\begin{abstract}
Following our previous work, we conclude that a GRB standard candle
constructed from the Ghirlanda et al. power-law relation between the
geometry-corrected energy ($E_{\gamma}$) and the peak of the
rest-frame prompt burst spectrum ($E_p$) is not yet cosmographically
useful, despite holding some potential advantages over SNe Ia.  This
is due largely to the small sample of $\sim20$ GRBs with the required
measured redshifts, jet-breaks, and peak energies, and to the strong
sensitivity of the goodness-of-fit of the power-law to input
assumptions.  The most important such finding concerns the sensitivity
to the generally unknown density (and density profile), of the
circumburst medium. Although the $E_p$--$E_{\gamma}$ relation is a
highly significant correlation over many cosmologies, until the sample
expands to include many low-$z$ events, it will be most sensitive to
$\Omega_M$ but essentially insensitive to $\Omega_{\Lambda}$ and $w$,
with some hope of constraining $dw/dt$ with high-$z$ GRB data alone.
The relation clearly represents a significant improvement in the
search for an empirical GRB standard candle, but is further hindered
by an unknown physical basis for the relation, the lack of a low-$z$
training set to calibrate the relation in a cosmology-independent way,
and several major potential systematic uncertainties and selection
effects.  Until these concerns are addressed, a larger sample is
acquired, and attempts are made to marginalize or perform Monte Carlo
simulations over the unknown density distribution, we urge caution
concerning claims of the utility of GRBs for cosmography and
especially the attempts to combine GRBs with SNe Ia.
\end{abstract}

\section{Motivations for a GRB Standard Candle}
It has long been recognized \cite{derm92,rutl95,cohe97}, that standard
candles constructed from long duration Gamma-ray bursts (GRBs) would
have several potential advantages over Type Ia Supernovae (SNe Ia),
the most important being high redshift detection.  Whereas detected
SNe Ia are currently spectrally classifiable out to a maximum of
$z\sim 1.7$ with {\it HST} \cite{ries01} (and in the future with {\it
SNAP} \cite{lind04}), $\sim25\%$ of GRBs with known $z$ ($10$ of $39$)
already have measured redshifts $>2$.  Although there are diminishing
returns for observations at higher redshifts --- which primarily probe
the matter-dominated regime --- such measurements may be of great
interest if the dark energy shows exotic time variation.  In practice,
$\sim50\%$ ($9$ of $19$) GRBs in the current sample of bursts with
measured redshifts ($z$), jet break times ($t_{\rm jet}$), and peak
energies ($E_p$), are in the redshift range $0.9 < z < 2$, which is
{\it already} comparable to the number of high-z SNe Ia discovered
with {\it HST} \cite{ries04a}.  This regime is clearly important for
constraining $\Omega_{\Lambda}$, $w$, its possible time variation, and
the transition redshift to the epoch of deceleration
\cite{ries04a,lind03}. In addition to high-$z$ detection,
$\gamma$--rays penetrate dust, GRB spectra \cite{band93} are simpler
than SNe Ia spectra, yielding potentially cleaner $k$-corrections
\cite{bloo01b}, and with massive star progenitors \cite{woos93}, any
long GRB evolution would likely be orthogonal to that of SNe Ia,
ensuring different systematics.  GRB standard candles could thus
provide a useful independent check to cosmography with SNe Ia.

\section{Present Status of the $E_p$--$E_{\gamma}$ Relation: Sensitivity to Input Assumptions}

Previous attempts to constrain cosmological parameters with GRB
energetics \cite{derm92,rutl95,cohe97,bloo03} were thwarted by what
are now known as wide distributions in the isotropic equivalent
energies ($E_{\rm iso}$) and beaming-corrected energies
($E_{\gamma}$), which span more than $\sim4$ and $\sim2$ orders of
magnitude, respectively \cite{frie05}.  In particular, the
once-promising $E_{\gamma}$ distribution \cite{frai01,bloo03}, widened
with the discovery of new low energy bursts (e.g. 030329). Expanding
upon the well known $E_p$--$E_{\rm iso}$ ``Amati'' relation
\cite{amat02}, Ghirlanda \etal \ recognized that many under-energetic
bursts appeared softer in the rest-frame prompt $\gamma$--ray spectrum
than those with higher $E_{\gamma}$, discovering the remarkable
$E_p$--$E_{\gamma}$ ``Ghirlanda'' relation \cite{ghir04}, which can be
cast as a power-law $E_p \propto (E_{\gamma})^{\eta}$.  In
\cite{frie05}, we confirm this correlation (Fig. 1a herein), and
demonstrate that although the goodness of fit to a power-law is
strongly sensitive to input assumptions (particularly the circumburst
density), the relation itself is still highly significant over a range
of plausible cosmologies.  As recently suggested
\cite{dai04,ghir04b,frie05}, the relation could be used to create a
standardized GRB candle with an empirical correction to the energetics
similar to the light-curve shape corrections used to standardize the
peak magnitudes of SNe Ia \cite{phil93}.  However, without knowing the
slope of the power law {\it a priori} from physics or from a low-$z$
training set, in the cosmographic context, it is imperative to re-fit
for the slope of the power law from the data for each cosmology
\cite{ghir04b,frie05} (Fig. 1a inset), lest circularity problems arise
(e.g. \cite{dai04}).  Even so, greater obstacles to cosmography with
current data involve small number statistics and the sensitivity of
the goodness of fit of the relation to input assumptions
\cite{frie05}.

\begin{figure}
\includegraphics[width=2.0in]{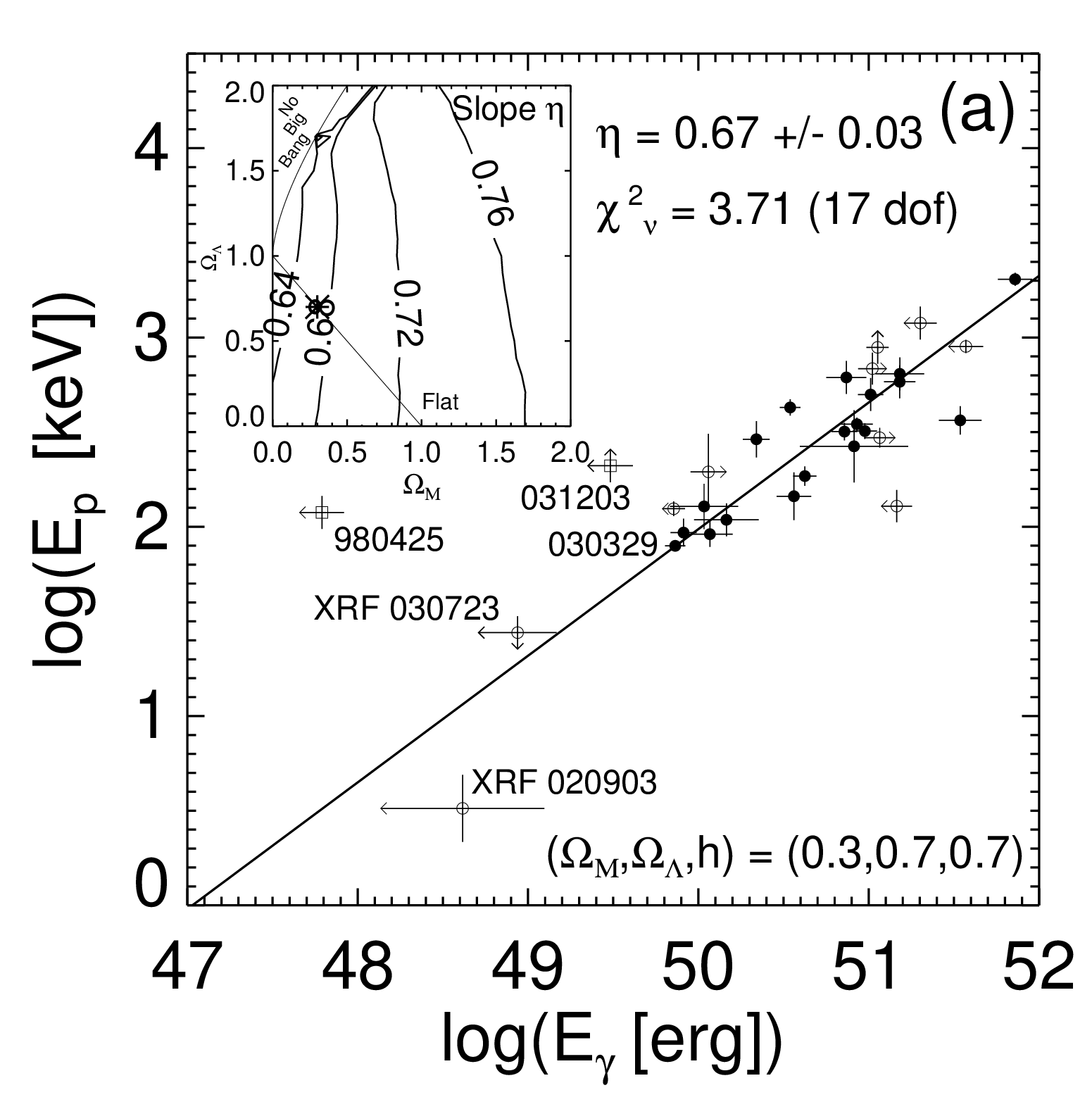}     % includes figure foo.eps
\includegraphics[width=3.2in]{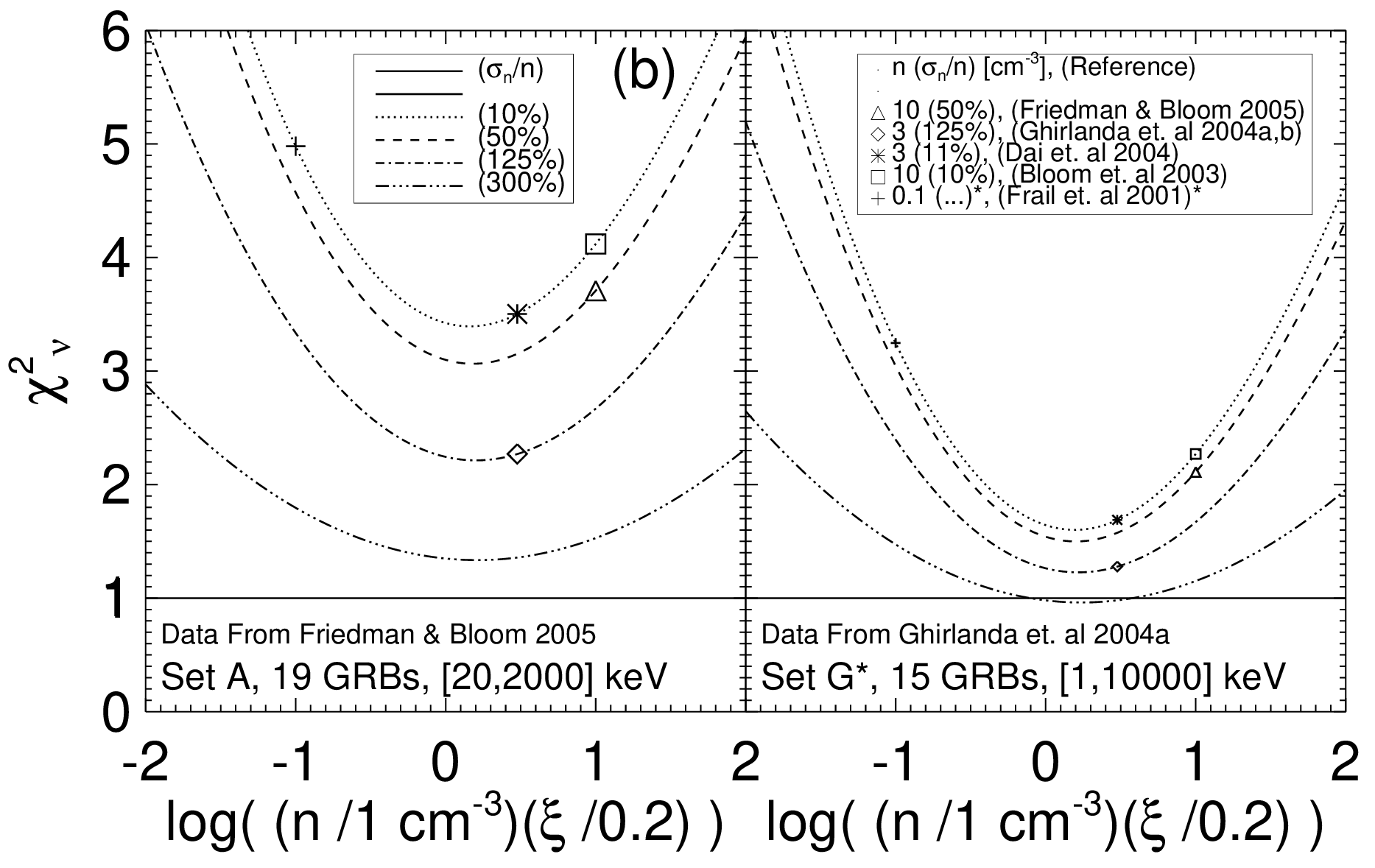}     % includes figure foo.eps
\caption{\scriptsize {\bf (a)} $E_p$--$E_{\gamma}$ relation, plotted from Tables 1,2 of \cite{frie05}, with only 19 GRBs used in the fit (filled circles). Upper/lower limits (open circles) are denoted with arrows.  Inset shows weak cosmology dependence of slope $\eta$.  {\bf (b)} Sensitivity of $\chi^2_{\nu}$ to assumptions for density $n$ (error $\sigma_{n}$). Data, are from \cite{frie05}: left, \cite{ghir04}: right. Curves are for $\sigma_{n}/n = [0.1,0.5,1.25,3.00]$ from top down.  Plot symbols show assumptions from previous work.  Independent of data set, $k$-cor bandpass, $n\sim 1$--$2$ cm $^{-3}$ minimizes $\chi^2_{\nu}$.  Derived $\chi^2_{\nu}$ are left: $3.71$ (17 dof), big triangle; right: $1.27$ (13 dof), small diamond.  In decreasing importance, discrepancy comes from data references, density assumptions, sample size, $k$-cor bandpass. {\bf (a)} and {\bf (b)} assume $\xi=0.2$, ($\Omega_M$,$\Omega_{\Lambda}$,$h$)=($0.7$,$0.3$,$0.7$).}
\end{figure}

Unlike peak magnitudes of SNe Ia light curves, computing $E_{\gamma}$
(see eq's 1, 2 of \cite{frie05}) depends on {\bf (a)} the cosmology,
{\bf (b)} a model assumption for the energy structure of the jet, {\bf
(c)} the effective rest-frame ``bolometric'' bandpass for the GRB
$k$-correction \cite{bloo01b}, and {\bf (d)} parameters which are
often unknown for most bursts; namely the ambient ISM density $n$
(possibly a stellar wind profile \cite{chev00}), and the $\gamma$--ray
creation efficiency ($\xi$), where the same values of $n$, $\xi$ are
assumed for all GRBs when unknown.  Assuming $n=10 \pm 5$ cm$^{-3}$,
$\xi=0.2$ (20\%), for our sample of 19 GRBs, we find a goodness of fit
of $\chi^2_{\nu} = \chi^2$/dof $=3.71$ (17 dof) for the $[20,2000]$
keV bandpass \cite{frie05}.  At this conference, Ghirlanda et al.
first reported $\chi^2_{\nu}=1.27$ (13 dof) for their sample of 15
GRBs \cite{ghir04c}.  In \cite{frie05}, we demonstrate that the
discrepancy in $\chi^2_{\nu}$ arises mainly from different references
for individual bursts (\idest \ small number statistics), and from
slightly different input assumptions for $n$, $\sigma_{n}$,
illustrated here in Fig. 1b.  Poor fits to the relation result in
unacceptable fits in the GRB Hubble diagram, rendering the $\chi^2$
contours over ($\Omega_M$,$\Omega_{\Lambda}$) meaningless; however,
this sensitivity also allows most outlier bursts in Fig. 1a to be made
consistent with the relation by changing the density (or its error) to
otherwise reasonable values \cite{frie05}.

Using the relation for cosmography, in \cite{dai04}, although the
authors claim tight constraints on the matter density (assuming
flatness), they fail to self-consistently re-fit the slope of the
relation for each cosmology and exclude several outlier bursts
(e.g. 990510, 030226) on grounds not adequately justified.  Recently,
those authors have improved upon these points in follow-up work
\cite{xu05}.  In \cite{ghir04b}, the authors re-fit the relation
self-consistently, but did not stress the cosmographic power of GRBs
alone, instead performing a joint fit with SNe Ia, claiming that the
joint fit is more consistent with flatness than SNe Ia alone.
However, the analysis in \cite{ghir04b} understresses the fact that
GRBs alone appear to favor a loitering cosmology \cite{frie05}
(although see \cite{firm05}).  Despite improvements, the most recent
follow up work \cite{firm05,xu05} does not address the sensitivity to
input assumptions \cite{frie05}.  However, by performing simulations
of future data \cite{xu05}, and developing new statistical techniques
\cite{firm05}, the recent work had indicated some potentially exciting
new directions for GRB cosmology.

\section{Future Prospects}
Although a major step forward, a GRB standard candle constructed from
the $E_p$--$E_{\gamma}$ relation can not yield meaningful constraints
on the cosmological parameters for the current data, mainly due to the
small sample (with very few low-$z$ events) and to the strong
sensitivity of the goodness of fit of the relation to density
assumptions.  Selection effects \cite{naka04,band05} and other
potential systematics \cite{frie05} must also be addressed before
using GRBs for precision cosmography. However, {\it Swift} should
detect $> 200$ GRBs over the next $\sim2$ years \cite{gehr04}.  Of
these, redshift constraints are expected for a majority of the bursts,
either from the on-board broad-band spectroscopy or ground-based
follow-up spectra. With early-time light curves from the {\it Swift}
UVOT instrument and a fleet of dozens of ground-based follow-up
programs, $t_{\rm jet}$ could be measured for a substantial fraction
of these bursts.  Unfortunately, future $E_p$ measurements may be
hindered by the relatively narrow spectral range of {\it Swift}
([15,150] keV), further strengthening the science case for the ongoing
symbiosis with {\it HETE II}, due to its larger [30,400] keV bandpass.

Even for a sample including an order of magnitude more GRBs with
measured $z$, $t_{\rm jet}$, and $E_p$, we believe that density
constraints will remain the limiting factor for cosmography, since
each requires detailed broadband afterglow modeling
(e.g. \cite{pana02}).  This is independent of a low-$z$ training set
or a theoretical prediction that constrains the slope of the relation
{\it a priori}.  Future work will then require using the known
information about the {\it distribution} of densities to marginalize
over or sample statistically from such a distribution with Monte Carlo
simulations \cite{frie05}.  While the current data do not uniquely
support a good fit for the relation to a power law, they certainly do
not rule out one.  As such, it is still possible that GRB standard
candles from this relation might place meaningful constraints on the
cosmological parameters, most notably the time variation of the dark
energy.

However, even if the relation is never able to seriously constrain
cosmology, with the beaming-corrected energy $E_{\gamma}$, it is more
physically motivated than the Amati relation \cite{amat02} (also see
\cite{naka04,band05}), it may lend insight into GRB radiation physics
\cite{ghir04,eich04,rees04}, and could help us to identify new classes
of GRBs (\idest \ different progenitors) \cite{frie05}.  Independent of
its ultimate fate as a potential GRB standard candle, the discovery of
the $E_p$--$E_{\gamma}$ Ghirlanda relation \cite{ghir04} clearly
represents an exciting new direction for the GRB field.

\acknowledgments
{\scriptsize A.S.F. acknowledges support from a National Science
Foundation Graduate Research Fellowship and the Harvard University
Dept. of Astronomy.  We thank K. Stanek \& P. Nutzman for reviewing
the manuscript. We thank G. Ghirlanda and collaborators for patience
in explaining to us the details of their analysis.}

\end{document}